\documentclass[referee]{aa}      
\usepackage{graphicx}
\usepackage{txfonts}
\usepackage{epsfig}
\begin{document}

\title{Gamma--Ray Bursts associated with Supernovae: 
A systematic analysis of BATSE GRB candidates}      
\author{Z. Bosnjak\inst{1}, A. Celotti\inst{1}, G. Ghirlanda\inst{2},
M. Della Valle\inst{3}, E. Pian\inst{4}}

\offprints{bosnjak@sissa.it}

\institute{$^1$S.I.S.S.A./I.S.A.S., via Beirut 2-4, I-34014 Trieste,
Italy\\ $^2$INAF-Observatory of Brera, via Bianchi 46, I-23807 Merate
(LC), Italy\\ $^3$INAF-Observatory of Arcetri, Largo E. Fermi 5,
I-50125 Firenze, Italy\\ $^4$INAF-Observatory of Trieste, via
G.B. Tiepolo 11, I-34131 Trieste, Italy\\}

\date{}
\titlerunning{GRBs associated with SNe}
\authorrunning{Z. Bosnjak  et al.}

\abstract{We examined the properties of a sample of BATSE Gamma--Ray
Bursts (GRBs) comprising events which have indications of association
with a supernova (SN), some on the basis of indications of
re--brightening in the optical afterglow light curve, but in most
cases based only on the `loose' temporal and directional coincidence
inferred from the cross correlation of catalogs.  Despite of the large
uncertainties in the latter selection method, the temporal and
spectral analysis reveal three interesting statistical results when
the sample is compared with that of all the BATSE GRBs: the GRBs
tentatively associated with SNe are found to predominantly (in $\sim$
80\% of the cases) have single-peaked light curves, a softer spectrum
(i.e. low energy power law index $\alpha \sim$ --1.5) and tend not to
follow the Lag-Luminosity and Isotropic Energy--Peak Energy
correlations.  These three independent statistical properties point
toward the existence of a significant number of under-luminous,
GRB~980425-like events constituting -- at least from an observational
point of view -- a tail or a separate class with respect to the whole
of the BATSE GRB events. The unusually high percentage of SN Ibc among
those identified by the catalog cross--correlation (factor $\sim 4$
higher than expected from SN catalog statistics) reinforces the
non-randomness of (some of) the selected events.  \keywords{Gamma
Rays:bursts -- methods:data analysis -- supernovae}
\titlerunning{Candidates GRB coincident with SN}}
\maketitle

\section{Introduction}

An increasing number of recent observations support the association of
(at least) some GRBs with SN events, stimulating both the theoretical
understanding/modeling of GRB progenitors and intensive observational
campaigns aimed at finding further and more detailed evidence of
associations.

GRB~980425 was the first burst with tantalizing evidence for a
physical association with a SN (SN~1998bw, Galama et al. 1998).  The
association of the two events has been discussed in the frame of
models involving the core collapse of massive stars (e.g.  Woosley et
al. 1999; MacFadyen \& Woosley 1999; Vietri \& Stella 1998).  However,
the unusual properties of both events within their classes -- i.e. the
high luminosity of the SN, the unusually low luminosity and
variability of the GRB $\gamma$--ray emission and the lack of detected
X--ray afterglow -- as well as the uniqueness of this association,
have not allowed to draw firm conclusions about the possible
progenitor for a few years.

More recently two further GRB/SN associations based on the
spectroscopic detection of a SN spectrum in the GRB afterglow added
robust evidence to the association of the two phenomena.  GRB~030329
(Stanek et al. 2003; Hjorth et al. 2003) has been associated to
SN~2003dh from the appearance of typical SN Ic spectral features in
its optical spectrum and the (under-energetic) GRB~031203 revealed an
optical spectrum consistent with the contribution of an underlying
type Ic SN (SN~2003lw; Malesani et al.  2004).

Despite of these promising findings, the paucity of cases of GRB
directly associated with SN Ib/c, some peculiar properties of both the
GRB and SN events in these few robust associations and the
uncertainties on the event rates in both classes (see however
Podsiadlowski et al. 2004) do not allow a clear insight and different
possibilities are still open: selection effects might play a role in
detecting SN features, there might be an (inverse) relation between
the energetics of the two events (under-luminous GRB vs Hypernova
event), or bursts associated with core collapse SN could represent a
subset of the whole population, which might be even characterized by
properties (e.g. energetics, duration, temporal variability)
intermediate between the two class of events (e.g. Della Valle 2004
for an overview; Soderberg 2004).
 
Given the limited number of `direct' associations (i.e. with evidence
of SN spectral features in the GRB afterglow), in order to test for
the nature of any physical connection and thus investigate the
progenitor(s) of GRB, other more indirect indications could be
considered to select candidates GRB/SN associations.  These include
the re--brightening in the optical afterglow light curve decay,
possible signature of SN emission emerging at times when the GRB has
sufficiently faded (Bloom et al.  1999; Germany et al.  2000).  Indeed
about ten such associations have been claimed on such a basis (Zeh et
al. 2004).  Interestingly Zeh et al. (2004) found indication of
re-brightening in all GRBs with known redshift $z<0.7$. Similar
evidence (i.e. a bump in the late time light curve) have been found
also in the case of a few X--ray Flashes (see Fynbo et al. 2004,
Soderberg et al. 2005). An even more indirect and uncertain, but
possibly more efficient approach, consists of systematic searches for
directional and temporal coincidences of GRBs and SN from the cross
correlation of GRBs and SN catalogs (Wang \& Wheeler 1998, Hudec et
al. 1999).  These searches are clearly affected by i) the large
uncertainties in the GRB position (typically of a few degrees) (e.g.
Kippen et al.  1998) leading to possible serious overestimates of the
number of real associations, but also ii) the completeness and flux
limit of the SN/GRB catalogs used.  Nevertheless, if the indications
by Zeh et al.  (2004) prove to hold, catalog cross-correlations should
indeed find evidence for a significant number of real associations.

In principle a more powerful method for selecting candidate
associations from the catalog cross correlation could be envisaged if
GRB associated to SN were characterized by particular temporal and/or
spectral properties with respect to the whole of the GRB
population. Indeed Norris et al.  (1999) explored the hypothesis that
there exists a subclass of GRBs related to SN whose $\gamma$--ray
prompt light curve presents a single smooth peak, similar to that of
GRB~980425 (as suggested by Bloom et al.  1998), but with negative
results. 

Here we focus on examining the possible peculiarities of the candidate
events by investigating the temporal and spectral properties of GRBs
for which there is an indication of being associated with a SN as a
``class''.  More specifically we considered direct and indirect
(tentative) associations, the latter ones based either on the light
curve re--brightening or on the positional-temporal coincidence from
catalogs cross correlations.  To this aim we searched the literature
for the most up-to-date sample of GRB/SN candidates including the
three robust cases of association (GRB 980425, GRB 030329, GRB 031203)
and the 11 cases where the afterglow light curve re--brightening gives
evidence for a coincident supernova explosion, that is GRB 021211
(Della Valle et al. 2003), GRB 011121 (Garnavich et al. 2003), GRB
020410 (Levan et al. 2004), GRB 041006 (Stanek et al. 2005) and 7 GRBs
examined by Zeh et al. (2004) (GRB 970228, GRB 980703, GRB 990712, GRB
991208, GRB 000911, GRB 010921, GRB 020405). From the available data
we find that the light curve of these events consists of a single peak
in 50\% of cases and that their spectral parameters indicate a soft
low energy spectrum. Given these suggestions we then extended the
number of candidates by including those from both past catalogue
cross--correlations and updating them using more recent GRB and SN
catalogs.

Clearly, if GRBs in the sample turned out to be statistically
(although not individually) characterized by any peculiar property --
with respect to a random GRB sample -- this would have interesting
implications: i) it would indicate the existence of a SN-related
sub--class of GRBs; ii) it would provide an empirical tool for
efficiently identifying such events; iii) most importantly, it would
offer clues on the physical understanding of their origin (indicating
a mechanism different from that underlying GRB without SN signatures).

The outline of the paper is the following.  The selected GRB sample is
described in Section 2 and its general properties are discussed in
Section 3.  In Sections 4 and 5 we present the detailed temporal and
spectral analysis.  We discuss in Section 6 the reliability of the
GRB/SN associations and the nature of the selected GRB by means of
(two of) the phenomenological intrinsic relations found for GRB with
estimated redshift.  Section 7 summarizes our conclusions.  We adopt a
$\Lambda$CDM cosmology with $\Omega_{\Lambda}=$0.7, $\Omega_{\rm
m}=$0.3, $H_{0}=$ 65 km s$^{-1}$.

\section{Sample selection}

Let us describe in detail the (heterogeneous) sample considered
here, which basically includes the BATSE GRBs with indications of
association with SN and for which temporal and spectral information
could be retrieved, for a total of 36 GRB/SN. The whole sample with
its basic properties is listed in Table~1 and for simplicity hereafter
referred to as the GRB/SN sample.  The three robust associations based
on the spectroscopic identification of a SN, namely GRB~980425,
GRB~031203 and GRB~030329 will be considered as reference cases
throughout this work.

More precisely the GRB/SN sample comprises:

\noindent 
i) 7 GRB/SN suggested by a bump in the late-time optical afterglow
(Bloom et al.  1999, Zeh et al. 2004), robust spatial and temporal
association (GRB~980425) or an indication of a line feature in X-rays
(Piro et al.  1999), or exceptional SN properties (high luminosity,
peculiar spectrum; e.g. Germany et al. 2000). Of all the claimed cases
we included those with available BATSE data;

\noindent
ii) 15 GRB/SN from the previous searches of temporal and spatial
coincidence based on the cross--correlation of BATSE and SN
catalogs: 11 pairs by Wang \& Wheeler (1998) and 4 pairs by Hudec et
al. (1999);

\noindent
iii) 14 GRB/SN identified in this work from an update of the catalog
cross--correlation (including approximately the dates previously not
covered, after Nov.  20, 1997 up to May 26, 2000).  In this period the
BATSE GRB
catalog\footnote{http://cossc.gsfc.nasa.gov/batse/BATSE$\_$Ctlg/index.html}
contains 721 GRB and the SN catalog of
IAUC\footnote{http://cfa-www.harvard.edu/iau/lists/SN.html} reports
495 SN.

The catalog cross correlation method that we adopted is based on the
following criteria. For the positional coincidence we required a SN to
be within the BATSE GRB error box to which the constant 1.6 degrees of
r.m.s.  systematic error has been added (see e.g. Meegan et al. 1996).
The exact SN phase at detection is not known (typically the optical
maximum is reached in $\sim$20 days) and furthermore no robust guess
is possible for the delay between the SN and GRB explosions, as any
estimate is strongly model dependent: $\sim$ a few months up to few
years are allowed in the supranova model (Vietri \& Stella 1998) while
the explosions occur almost simultaneously in the collapsar model
(e.g.  MacFadyen \& Woosley 1998).  Although the latter case is
theoretically and observationally favored (e.g. Della Valle 2004 for a
review), a large temporal window has been allowed for the possible
correlation of the two events, requiring the SN date to be between -10
to +30 days with respect to the GRB trigger.  No further constraints
were imposed in the selection of either events and thus the results
are biased by the SN sky distribution and the detection incompleteness
at faint magnitudes; we considered both long and short GRB and no
limit was imposed on the GRB count rate. Also, despite of the
indications from the three robust associations, we did not impose any
selection on the SN type in our search (see however Section~5).

Note that in the previous catalogs cross correlations analysis
different search conditions were imposed.  Wang \& Wheeler (1998)
searched for SN with coordinates within 2$\sigma$ BATSE error boxes
and supposed different time delays (according to the SN type) between
SN and GRB.  They also limited their search to SN at $z <$ 0.1. As
this limit, however, would exclude 2 (GRB~030329 and GRB~031203) out
of the 3 more robust associations we did not impose any redshift cut.
Hudec et al.  (1999) reported the correlations for SN detected within
the GRB error box up to 30 days after the GRB. Thus the compiled
sample is not homogeneous as different criteria were applied. We
considered re-examining the associations for the periods examined by
Wang \& Wheeler (1998) and Hudec et al. (1999). However the BATSE
instrumental trigger parameters have been changed during the BATSE 9
yr activity (Paciesas et al. 1999), until February 1997, when they
have been fixed.  Therefore, a re-analysis would have been affected by
this further bias, and thus not led to any increased completeness of
the GRB/SN sample.

This updated cross-correlation yields 16 new GRB/SN coincident events.
However, coordinates from the
IPN\footnote{http://www.ssl.berkeley.edu/ipn3/interpla.html} for two
GRB revealed that the associations were spurious (and thus excluded
from the sample). In Fig.~1 error boxes and time delays are displayed
for 31 GRB/SN (2 GRBs had multiple associations and 3 GRBs had an
indication of an underlying SN from the afterglow observations).

The significance of such blind (catalog based) searches was studied by
Kippen et al.  (1998), who assigned a significance to the observed
correlation statistics by generating a set of random GRB catalogs.
They find no evidence for association for GRBs having precise
BATSE/Ulysses localizations, although they could not exclude it for a
subset of weak bursts.  While these results point toward a low
chance to select real coincidences with such a procedure, the
peculiarity of some properties of the GRB/SN sample (shown below) is
instead suggesting that such method does not select a random sub-sample
of BATSE GRBs.

\begin{figure}
\centerline{\epsfig{figure=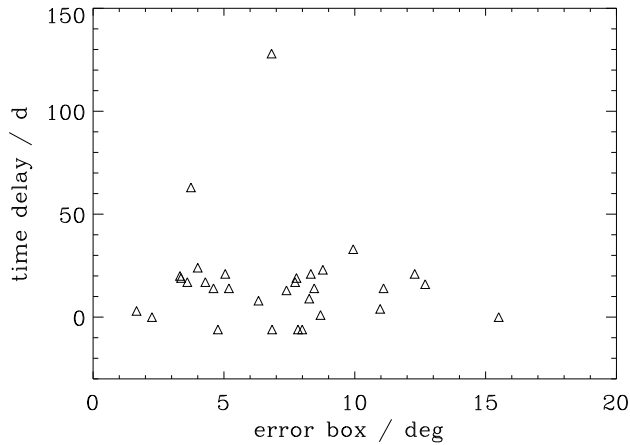,width=9.2cm}}
\caption{BATSE error boxes and time delays for the GRB/SN sample. The
time delay is between the SN discovery date and the GRB event.}
\end{figure}
    
\section{GRB/SN sample properties}

Based on the properties of GRB~980425 it has been suggested (see e.g.
Bloom et al.  1998) that, if there is a physical connection between a
sub-sample of GRBs and SN, such GRBs might exhibit distinct
characteristics.  Therefore we have compared the global properties of
the GRB/SN sample with those of the entire BATSE sample.

No qualitative (nor statistical) differences are found as far as
duration (represented by $T_{90}$), spectral hardness (HR32) ratio and
total burst fluence are concerned (see Fig.~2, top and bottom panels),
as both the range and shape of the GRB/SN distributions are
indistinguishable from those of the rest of GRBs in the BATSE catalog.

As expected from the typical worsening of the positional accuracy with
decreasing photon statistics (see Kippen et al. 1998), the GRBs
identified by the catalogs cross--correlations are mostly weak,
with typical peak (50-300 keV) flux $<$ 1.5 phot cm$^{-2}$ s$^{-1}$
(see Fig.~2).

While no peculiarity is found for the above quantities in the GRB/SN
sample, in the next Section we present a more detailed analysis of
their GRB temporal profile and spectra\footnote{Note that in the
following the GRB/SN events reported in each case/figure is different,
depending on the quality of the available data.}. Notice that the bias
on the flux as well as the selection criteria of the GRB/SN sample
result in a ``non optimal'' statistics for the temporal and spectral
analysis. This in turn requires a careful selection of a BATSE GRB
``control sample'', as clearly previous analysis of BATSE GRBs have
been typically performed on brighter samples.

\begin{figure}
\centerline{\epsfig{figure=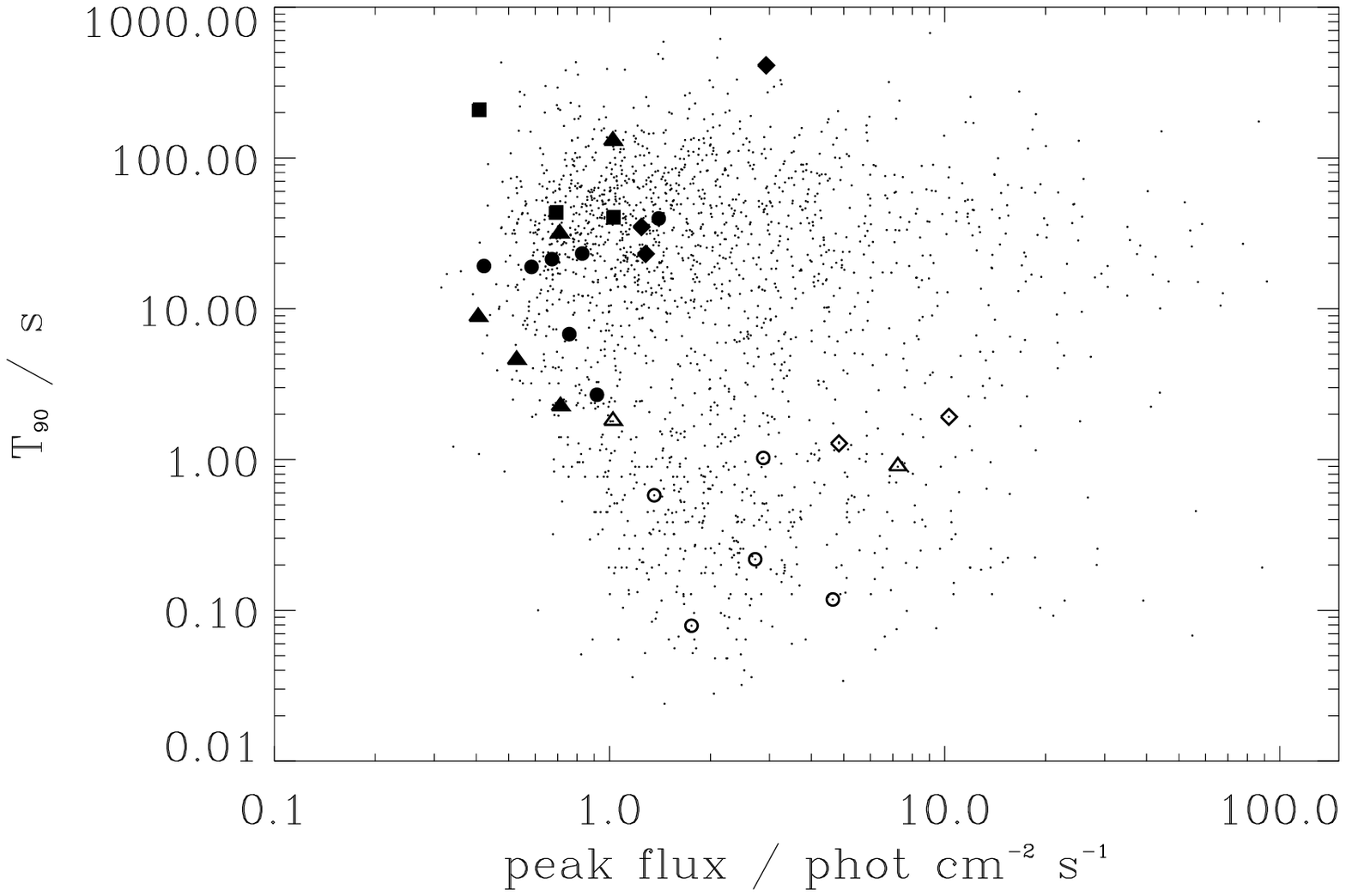,width=9.2cm}}
\centerline{\epsfig{figure=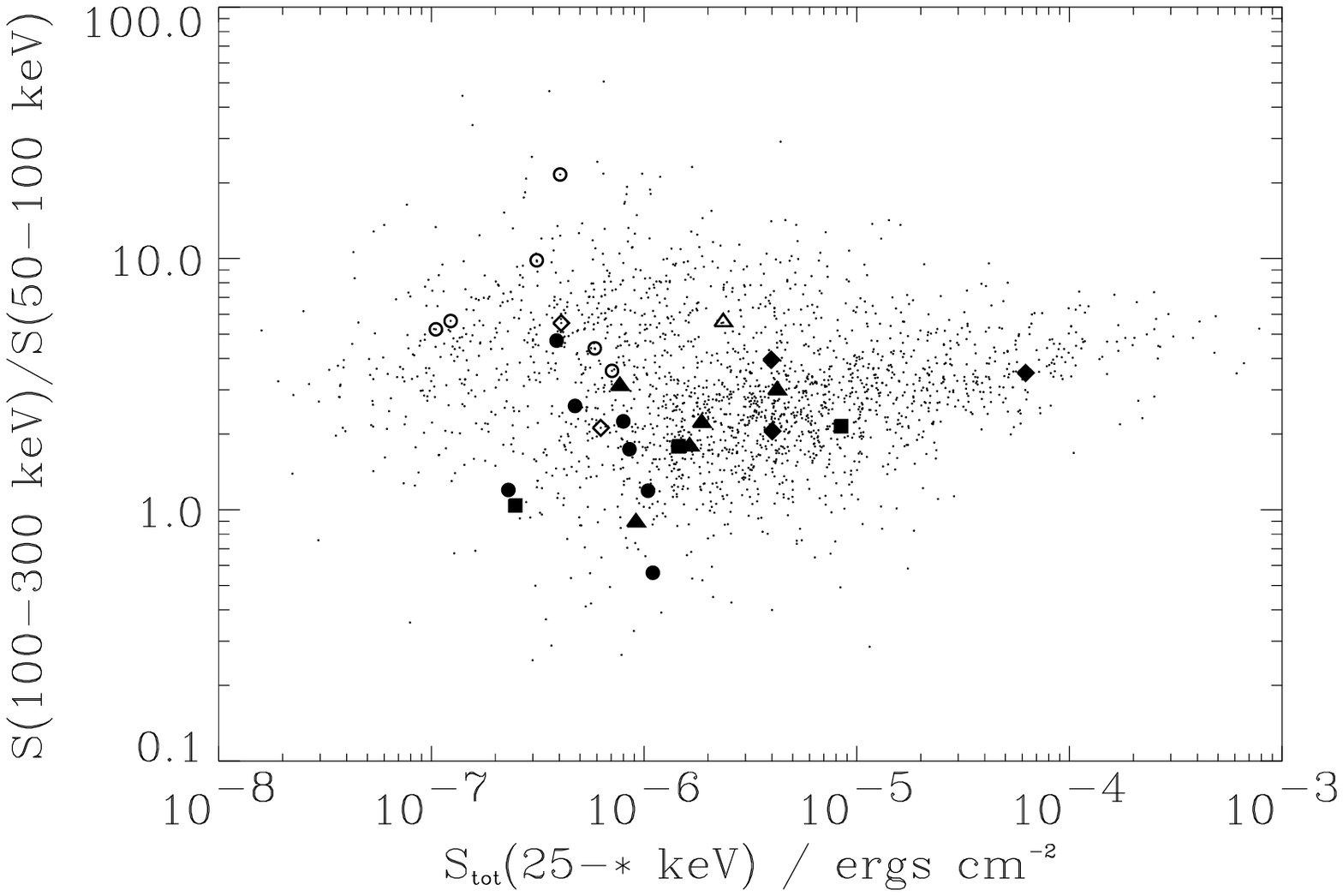,width=9.2cm}}
\caption{Comparison of the durations $T_{90}$, peak fluxes, hardness
ratios (i.e. HR32) and total fluences of the GRB/SN and the entire
BATSE samples.  Triangles, squares and circles indicate the tentative
associations found by Wang \& Wheeler (1998), Hudec et al. (1999) and
this work, respectively.  Diamonds are the GRB with an indication of a
SN event in the afterglow.  Filled symbols represent long GRB
($T_{90}>$2 s).}
\end{figure}

\section{Light curves analysis}

In this Section we present the analysis of the temporal profiles of
the GRB/SN sample and compare the results with the (known) properties
of larger GRB control samples.

While the temporal profiles of GRBs display a wide diversity (see e.g.
Norris et al. 1999) we recall for future reference some of the
important systematic findings on their properties: (1) flux peaks
occur later at lower energies (spectral lags - Norris et al. 1996,
Norris et al. 2000); (2) the pulse parameter distributions are
consistent with lognormal distributions (Quilligan et al.  2002); (3)
less variable GRBs are found to be intrinsically less luminous
(variability-luminosity correlation, Fenimore \& Ramirez-Ruiz 2000).

\subsection{Analysis}

For most of the GRBs (16/36) we analyzed the time profiles from the
BATSE concatenated 64-ms data derived from the eight BATSE Large Area
Detectors (LADs); these data represent the detector count rates in 4
energy channels (25-50, 50-100, 100-300 and $>$300 keV).  We fitted
both the energy--integrated flux and the individual 4 energy channel
light curves, according to the count statistics available for each
event.  The duration of 6 events was shorter than 2 s: in these cases
we could thus analyze the Time-Tagged Event (TTE) data with 2-ms time
resolution (which are buffer--size
limited\footnote{http://cossc.gsfc.nasa.gov/batse/batseburst/tte/index.html})
integrated over the 4 energy channels.  In the analysis of these data
type we also accounted for the $\approx$30 ms offset between trigger
time (as reported in the BATSE catalog) and TTE data load time.  For
the remaining 14 GRB/SN events either no reprocessed light curves were
found in the public archive (9 cases) or the low statistics of the
light curve did not allow any reasonable fit (5 cases).

First we determined for every GRB the background count rate selecting
two time intervals not containing the
burst\footnote{http://cossc.gsfc.nasa.gov/batse/batseburst/sixtyfour$\_$ms/
bckgnd$\_$fits.html}, and modeled it with a 2nd order polynomial.  We
then built a pulse model by adding the best fit background model to
the lognormal pulse profile (Brock et al. 1994) and fit the light
curve as:

\begin{displaymath}
I(t)= \left\{ \begin{array}{ll} A_{\max} ~
exp\left[{-\frac{1}{2}{\Big(
\frac{\log(t-t_{0})-\mu}{\sigma}\Big)}^{2}}\right], & t>t_{0} \\ 0 &
\textrm{otherwise}
\end{array} \right.
\end{displaymath}  
where $\mu$ and $\sigma$ are the mean and standard deviation,
respectively, and the time is scaled to the pulse onset $t_0$
\footnote{The lognormal pulse model is skewed to the right (as
determined by $\sigma$).}.  Two reasons are behind this choice of
profile: 1) most of the GRBs in the sample appear as single-peaked (on
the 64 ms and 2 ms timescale for long and short bursts, respectively);
2) it allows a direct comparison with the results of previous studies
(e.g. Norris et al. 1999).  The best fits ($\chi^2\sim$ 0.83 -- 1.18)
are obtained for the integrated light curves, and in 11 (out of total
22) cases an acceptable fit was obtained for individual channel light
curves.

\subsection{Results}

The resulting distribution of $\sigma$ is shown in Fig.~3 (top panel)
for all the pulses in the light curves of the 22 GRBs for which the
fit was performed. The distribution of $\sigma$ vs.  $\mu$ is reported
in the bottom panel.  As expected short bursts are characterized by
$\sigma$ lower than long ones and their pulses are less asymmetric.
While the statistics is rather poor, we note a global tendency for an
anti--correlation between $\sigma$ and $\mu$ both in long and short
GRB/SN: there is no pulse with large $\mu$ (i.e. long rise time) and
large $\sigma$ (i.e. very asymmetric), namely they are typically
FRED-like (i.e. Fast Rise Exponential Decay).

\begin{figure}
\centerline{\epsfig{figure=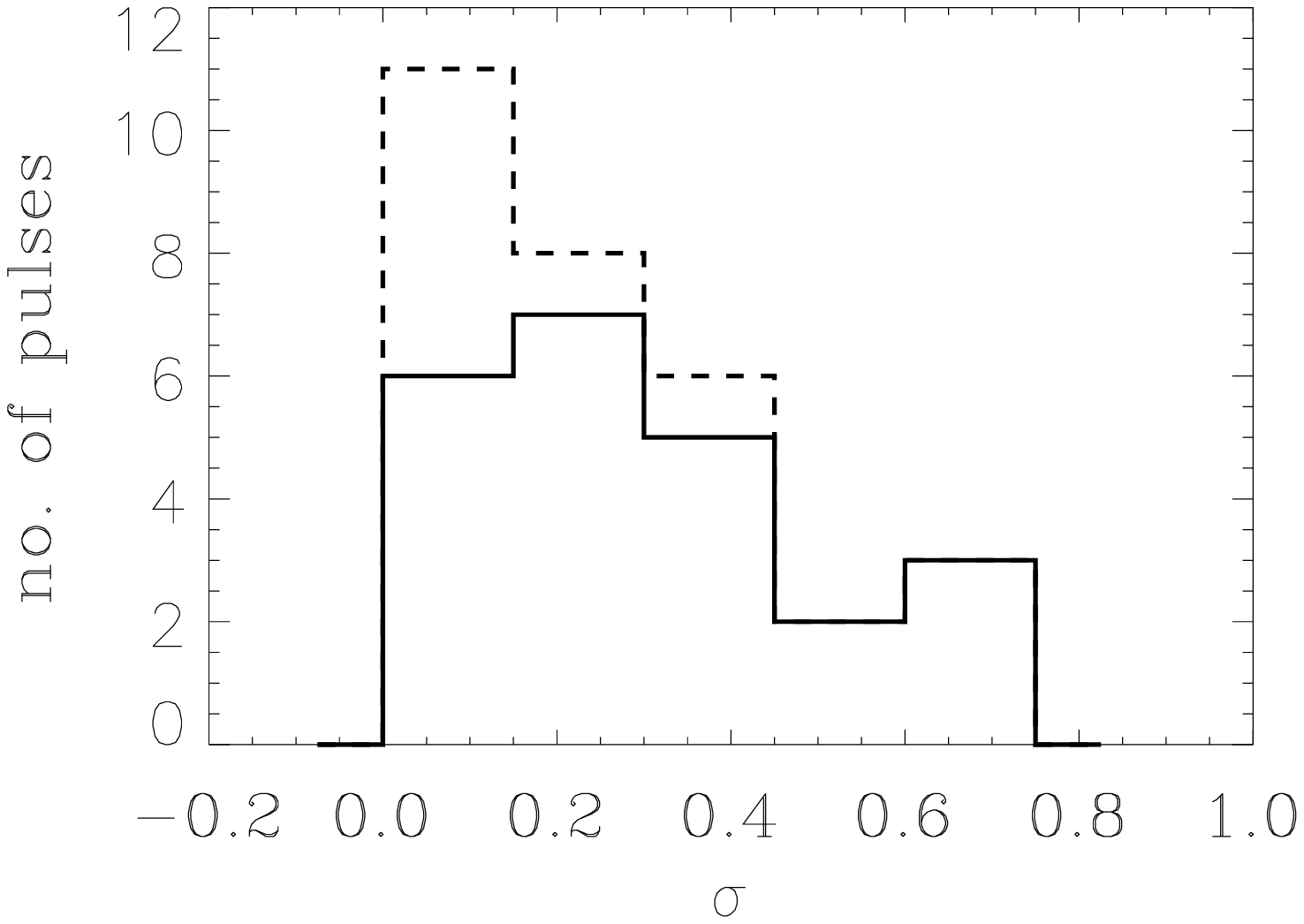,width=8.2cm}}
\centerline{\epsfig{figure=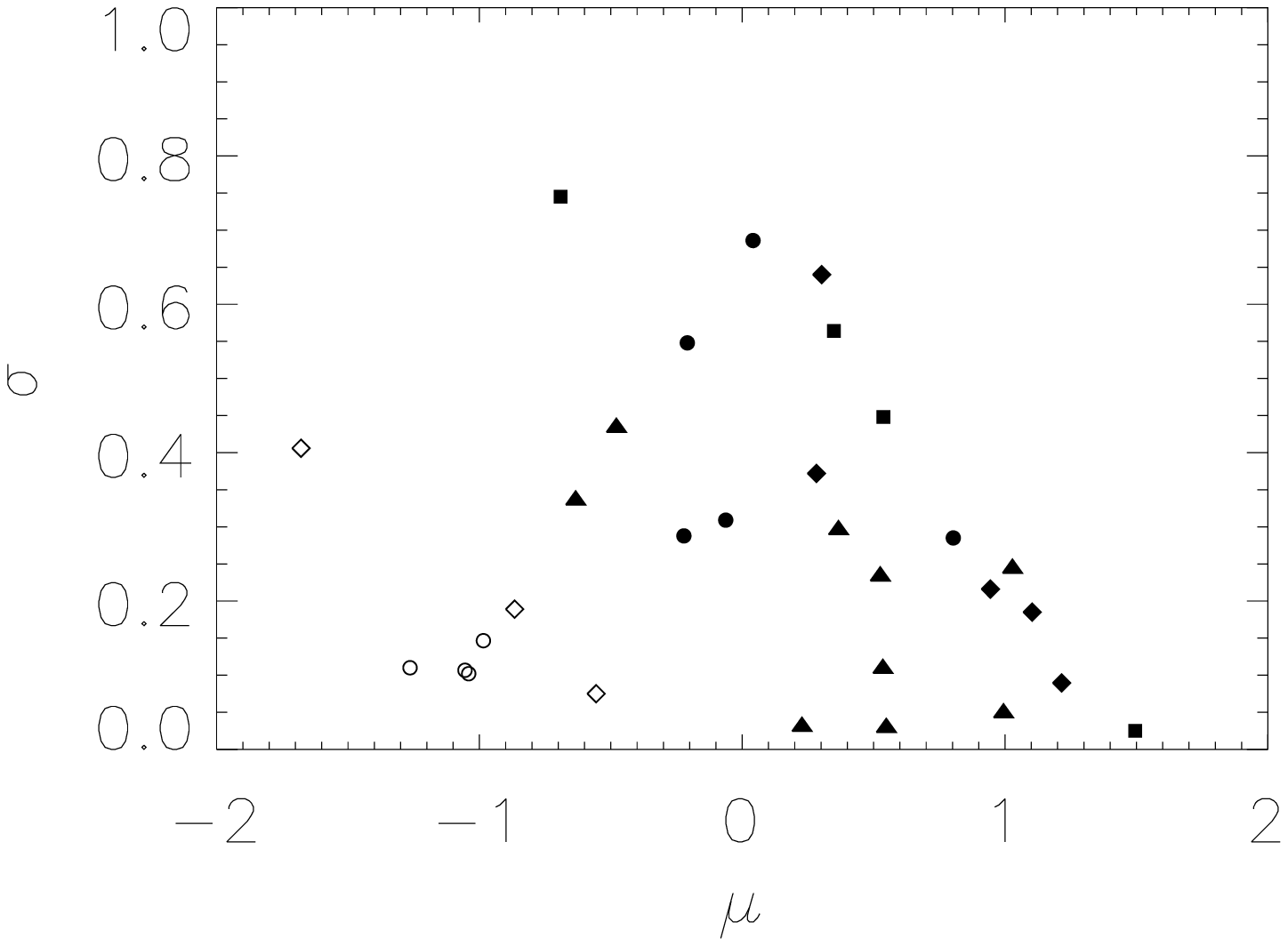,width=8.2cm}}
\caption{Top: Distribution of $\sigma$ for the lognormal pulse model.
The solid and dashed lines represent the parameters for long bursts
and the contribution of short bursts. All the pulses in the time
history are represented. Bottom: $\sigma$ vs $\mu$, i.e. the mean in
log time. Symbols as in Fig.~2.}
\end{figure}

\begin{figure}
\centerline{\epsfig{figure=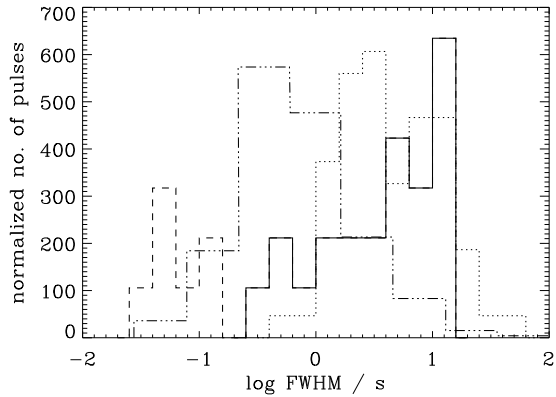,width=8.2cm}}
\caption{Distributions of full widths at half maximum.  The
solid/dashed lines represent the contribution of long/short GRBs.  The
dotted line refers to the single-peaked bursts analyzed by Norris et
al (1999), while the dash-dotted line are the bright, long bursts
analyzed by Quilligan et al (2002). For clarity the distributions are
normalized to the maximum number of data points (i.e. long bursts of
Quilligan et al 2002).}.
\end{figure}

In order to compare the GRB/SN pulse properties with a control sample
we examined also the distribution of their FWHM (64 ms time
resolution).  In Fig.~4 this is shown together with the distributions
derived by Norris et al. (1999) and Quilligan et al. (2002).  The
latter, referring to the brightest (long) 319 BATSE GRBs, is very
broad, peaking around 0.7 s, and significantly different from that of
the GRB/SN sample. Interestingly, however, the FWHM distribution for
the subset of long GRB/SN events (solid line in Fig.~4) appears to be
similar to that derived by Norris et al. (1999) from the analysis of a
sample of long, single-peaked GRB. Indeed a K-S test
\footnote{While formally the results by Norris et al (1999) refer to
  fits of count rates from channels CH2+3, typically the signals in
  CH1 and CH4 are weaker, thus not significantly affecting the
  comparison.}  returns (for long, single-peaked GRB/SN) a probability
  P$_{KS}$=0.98 that the two data sets originate from the same parent
  distribution.

In Fig.~5 (top panel) the pulse amplitudes (only for 64 ms resolution
timescale) vs.  FWHM are displayed.  Quilligan et al.  (2002) find a
weak negative correlation between these parameters. While there is no
evidence for a trend for the whole GRB/SN sample, an anti-correlation
(Spearman rank correlation coefficient of -0.65 and probability
0.003) is present among the GRB/SN selected on the basis of the
cross-correlation only.  The interpretation of this is not clear: it
might be simply due to the bias toward low fluxes in GRB/SN selected
from the catalog cross--correlation and/or indicate that indeed the
latter are largely spurious associations, not representing (atypical)
properties of GRB physically associated with SN events.

A tendency for wider pulses to have larger time lags appears instead
for the GRB/SN sample (Fig.~5, bottom panel). There the spectral lag
(i.e. the delay between the peaks of the low with respect to the high
energy channel) is plotted against the FWHM of the fitted pulse.
This behavior is consistent with what reported by Norris et al. (1996)
on the basis of a sample of 46 bright, long bursts: they found that
the lag results mostly from the shift in the pulse centroid and thus
the correlation is induced by the narrowing of the pulses toward
higher energies.  Different interpretations have been proposed for such
property (e.g. Salmonson 2000, Schaefer 2004) and very recently it has
been suggested that it might be accounted for in the jet-viewing-angle
scenario (Ioka \& Nakamura 2001) where both (small) pulse amplitudes
and (long) spectral lags are related to (large) viewing angles of the
jet.  As shown in the top and bottom panels of Fig.~5, in the case of
the GRB/SN sample (but excluding the 7 GRB/SN determined by the light
curve bumps) small amplitude pulses could correspond to large spectral
lags.

\begin{figure}
\centerline{\epsfig{figure=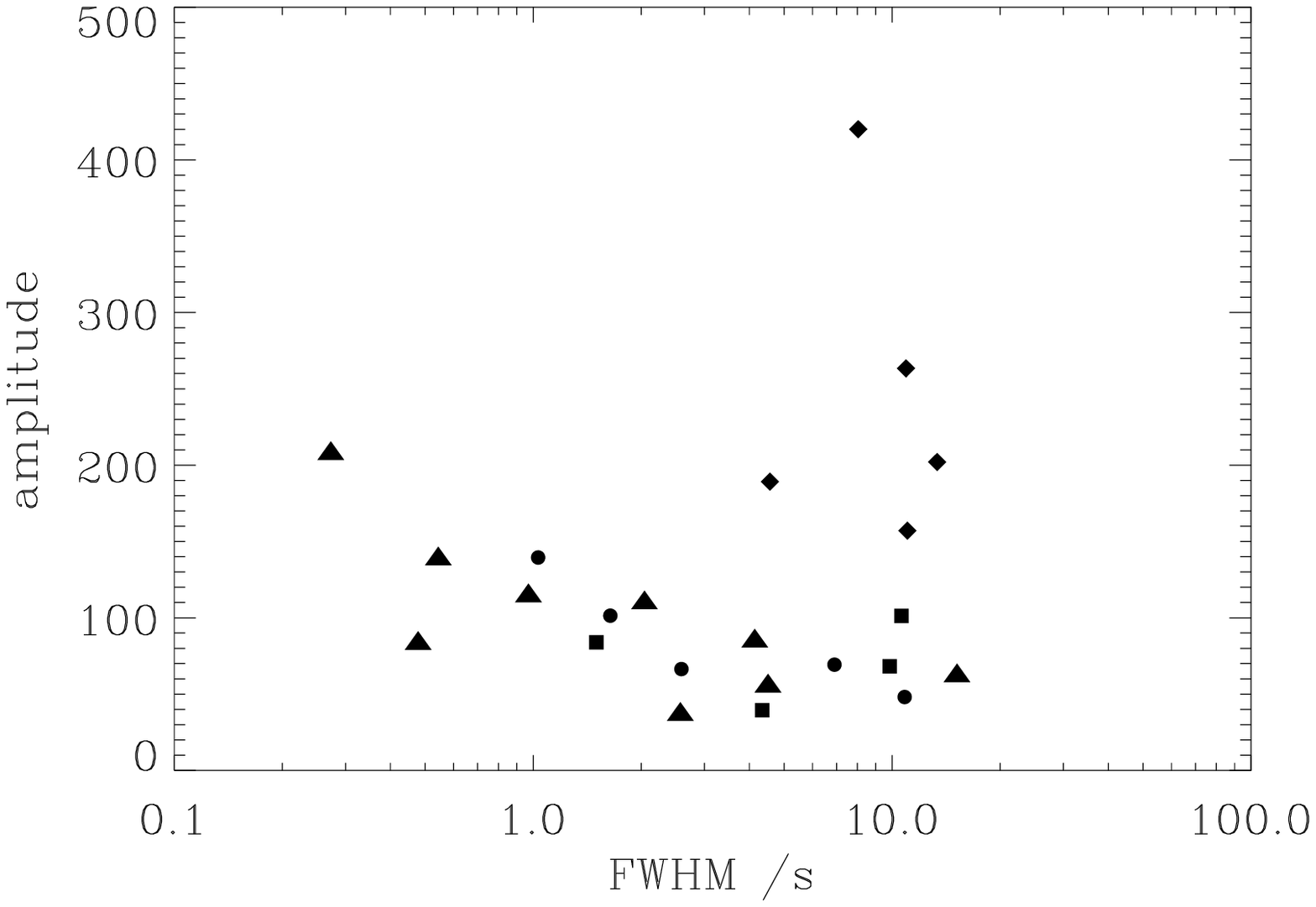,width=8.2cm}}
\centerline{\epsfig{figure=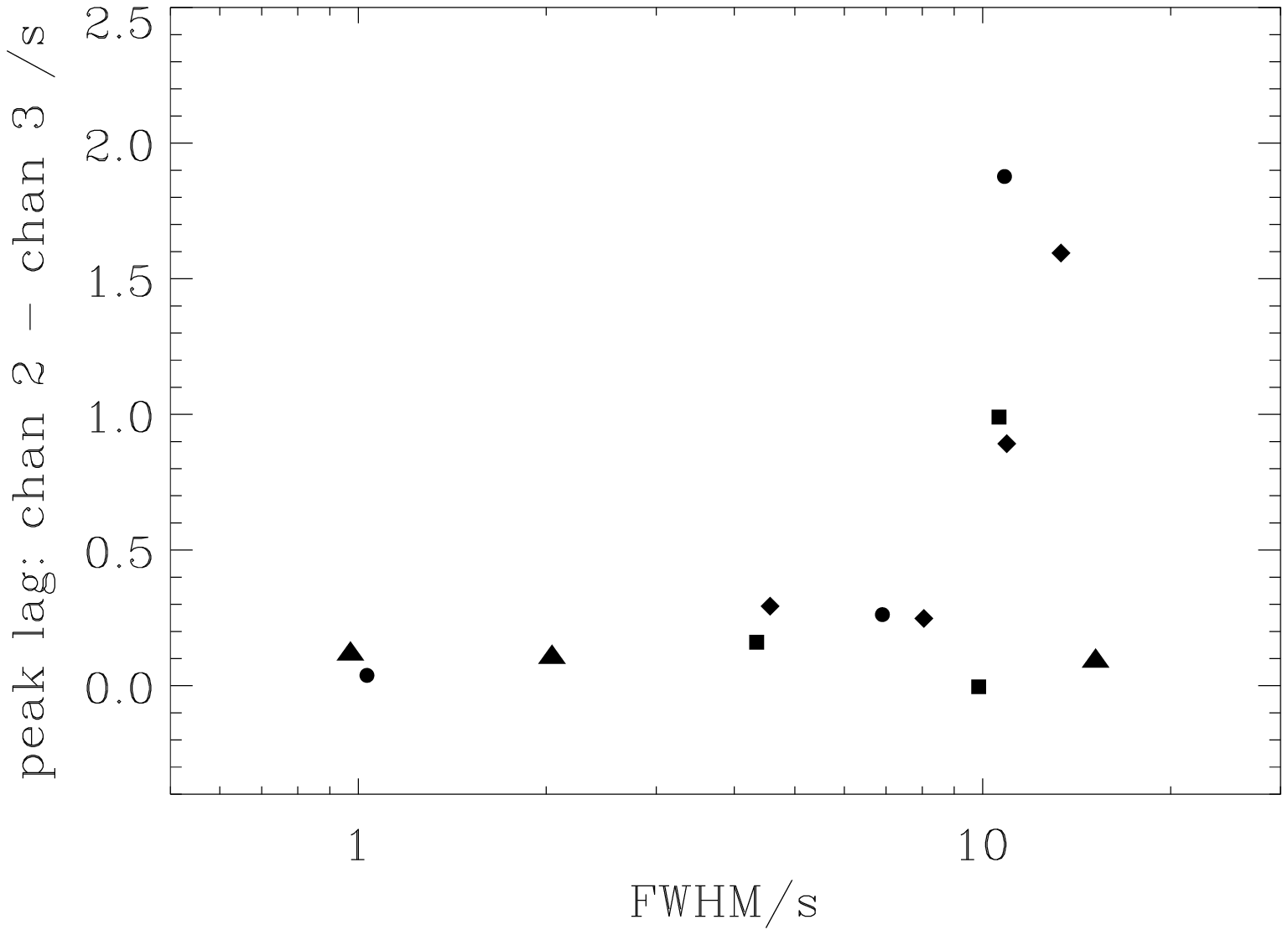,width=8.2cm}}
\caption{Top: Amplitude (fitted at 64 ms resolution) vs. FWHM as
obtained from the fit to the total counts in the four channels.
Bottom: Time lag between fitted peaks (in CH2 and CH3) vs. FWHM. Only
long bursts are shown. Symbols as in Fig.~2.}
\end{figure}
 
To summarize: within the limited statistics and differences in the
selection criteria of the GRB/SN sample, the GRB/SN events appear to
be broadly similar to the whole of the BATSE GRB population.  However,
the analysis of the FWHM distributions highlighted one intriguing
possibility, namely that the GRB/SN events behave more typically like
single--peaked GRBs. Although aware that the bias toward weak GRB in
the sample might be responsible for the above finding -- Norris et
al. (1999) already cautioned that dim bursts might have a multiple
pulse structure which is difficult to detect -- we explored this
aspect in detail.  

\subsection{A connection between single-peaked GRBs and SN?}

As already mentioned, Bloom et al.  (1998) suggested that GRBs
associated with SN might have distinctive temporal and spectral
properties, similar to those of GRB~980425, namely a smooth, single
peaked, long-duration ($T_{90}$=34.8 s) light curve and faint emission
above 300 keV.  Norris et al.  (1999) considered this issue and
searched the BATSE catalog with these criteria, but no other event
similar to GRB~980425 and spatially/temporally correlated with any
known SN was found. In our GRB/SN sample we instead found that the
large majority ($\sim$ 80 \%) of GRB are single peaked. Of those not
selected by the catalog cross--correlation 5 out of 7 GRBs are
single-peaked events.  It is worth stressing that of the other two
spectral associations, GRB~031203 shows a single peaked light curve
(Soderberg et al. 2004), while GRB~030329 presents different temporal
properties, i.e. a profile with two peaks of similar intensity
(Vanderspek et al. 2004).

In order to quantitatively test whether the GRB/SN sample comprises an
unusually high number of single--peaked GRBs (1) it has been necessary
to determine an objective way of characterizing the
``single-peakness'' and (2) due to the limited size of the GRB/SN
sample (29 GRBs with 64 ms data) the probability of finding a similar
large percentage of single--peaked events had to be computed via
random extractions of GRB sub-samples from the BATSE catalog.
 
\subsection{Method}

We thus built an algorithm that identifies single--peaked bursts in
the BATSE catalog.  As the time profiles of GRBs differ to great
extent it is difficult to have a unique approach applicable to every
burst: eventually we imposed conditions just to determine whether the
burst has one or more than one peak.

Li \& Fenimore (1996) proposed a method for identifying peaks in GRB
time histories, and a modification of this was also applied by Nakar
\& Piran (2002) in the temporal analysis of short bursts.  It consists
of defining a count bin $C_{\rm p}$ at time $t_{\rm p}$ as a candidate
peak if it satisfied the condition:
\begin{equation}
C_{\rm p} - C_{1,2} \ge N_{\rm var} \sqrt{C_{\rm p}},
\end{equation}      
where $C_{1,2}$ are neighboring count bins at times $t_{1,2}$.
$N_{\rm var}$ determines how rigorous the search is: too large $N_{\rm
var}$ might clearly result in an algorithm insensitive to faint peaks;
too low $N_{\rm var}$ could identify statistical variations as
peaks. The previous works showed that the choice of $N_{\rm var}$=5
optimizes the algorithm.

The search was performed on the background-subtracted light curves,
and thus a linear background was fitted to data before and after the
burst trigger, and subtracted.  The resulting time histories were
binned to 128 ms resolution.

As one of the problems encountered was that sharp structures in the
time history could be mis-identified as peaks, we applied a wavelet
analysis for determining the dominant modes of variability.  Wavelet
transform is convenient to use when the analyzed signal consists of
short spikes, that are not well approximated when doing a Fourier
analysis: the wavelet coefficients resulting from the convolution of
the signal with the wavelet basis functions represent the signal
variation at a particular resolution (see e.g. Kolaczyk 1997,
Quilligan et al.  2002 for details). We reconstructed the original
signal by keeping only the maxima of the wavelet transforms; a very
limited number ($\sim 25$) of the wavelet coefficients was used to
reproduce the signal. Although this does not guarantee the generation
of a smooth light curve, with this approach the unique peak finding
condition could be applied for the whole set of GRBs.

The search for peaks was limited to the burst region, excluding
periods before the burst which may contain a precursor.  We tested
several choices for the condition of peak finding, optimizing it to
recognize what would be visually identified as a single-peaked GRB,
and implemented the slightly modified peak-finding conditions: a)
adopted the value $N_{\rm var}$=7, and b) added the condition for a
peak to have at least 20 \% more flux than the neighboring 'valleys'.

\subsection{Results}

\begin{figure}
\centerline{\epsfig{figure=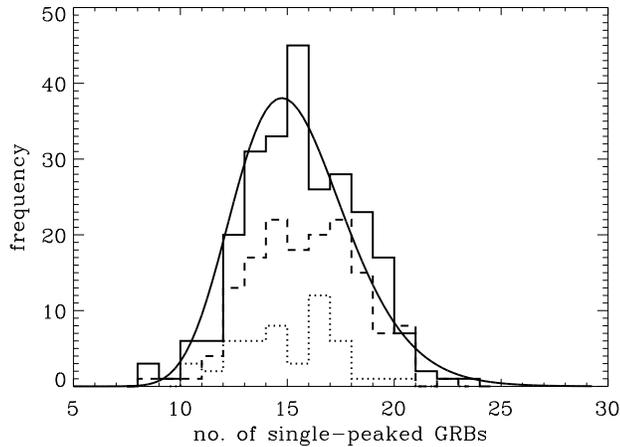,width=9.2cm}}
\caption{Distributions of the number of single--peaked GRB in
sub-samples of $\cal{N}$=29 GRB. The dotted, dashed and solid
histograms represent the distributions for respectively $n$=50, 150,
250 extractions out of the total sample of 1445 bursts.  The lognormal
fit corresponds to the case $n$=250.}
\end{figure}

We applied the algorithm to the GRB/SN sample and this defined 23 (out
of 29) single--peak bursts: this differs by less than 10\% from the
number we find by visual inspection.

To establish the chance probability of such occurrence, we selected
GRBs from the BATSE catalog having fluences in the range span by the
GRB/SN events (between $6.9\times 10^{-8}$ and $8.4\times 10^{-6}$ erg
cm$^{-2}$), ending up with 1587 GRBs, and systematically applied the
algorithm to their 64-ms resolution data (for 142 out of them the
algorithm did not identify any peak, as a consequence of the rigorous
conditions imposed).  We then randomly extracted sub-samples of the
same size ($\cal{N}$=29) of the GRB/SN sample, repeating the
extraction $n$ times. In Fig.~6 we show the distribution of the
fraction of single--peak GRBs for $n$=50, 150 and 250.

We fitted a lognormal function to the probability
distribution\footnote{Lognormal distributions have been already used
to describe the distributions of various GRB properties (see e.g.  Li
\& Fenimore 1996, Quilligan et al. 2002)}, i.e:

\begin{equation}
P(x)=\frac{1}{\sqrt{2\pi}\sigma x}e^{(-{(\ln(x)-\mu)}^{2}/2\sigma^2)},
\end{equation}   
for the case $n=$250 and find a 0.4 \% probability of having 23/29
single-peaked bursts.  If the number of single--peaked bursts in the
sample were 10\% lower (i.e. 20 or more events) the probability of
observing it would still be $\sim$ 5\%. Note that the distribution has
a maximum for $\sim$15 single--peaked bursts; this is consistent with
the result by Norris et al. (1999) who found 68/116 bursts with at
least one pulse -- suggesting however that it is difficult to assign
an exact number of pulses as they observed a continuum of lower
emissions superposed on otherwise single--peaked events.

The homogeneous treatment of the GRB/SN and random control sub-samples
should ensure that the above result is unbiased against effects that
may vary the number of detected pulses in the GRB time history, such
as the rigorous conditions on peak candidates, the fluence range, the
(still present) spikes mis-interpreted as pulses and the presence of
short GRBs which in fact appear to be multi-peaked when seen at 2 ms
resolution (there was 23\% of short bursts in the sample).

\section{Spectral analysis}

Let us now consider whether GRB/SN show any peculiarity in terms of
spectral properties.  We recall that the spectrum of GRB~980425 is
`unusual', peaking at $\sim$ 100 keV (Jimenez et al. 2001), below the
typical 300 keV peak energy. Also GRB~030329 is characterized by a
similar low peak energy (Vanderspeck et al. 2004), while GRB~031203
has a larger peak energy ($>$ 212 keV, Sazonov et al. 2004).

GRB spectra in the BATSE energy range (25-1800 keV) are usually
described via the phenomenological Band representation (Band et
al. 1993), which consists of two smoothly connected low- and high-
energy power laws.  Despite of a non--negligible dispersion of the
fitted parameters, the typical values for the low and high energy
power-laws photon indices are $\alpha\sim -1$ and $\beta\sim -2.5$,
respectively, and the spectral peak energy $E_{\rm peak}$ (i.e. the
peak in a $\nu F\nu$ representation) is around a few hundred keV
(Preece et al.  2000).  The Band model in general successfully
represents both the time integrated as well as the time resolved
spectra (Ford et al. 1995).

Following the widely adopted methodology described by Preece et al.
(2000), we analyzed the GRB/SN spectra obtained by integrating the
signal over the burst duration (a time resolved spectral analysis was
possible only in a few cases due to the low signal--to--noise ratio).

\subsection{Analysis}

The spectral analysis was heavily limited because in many cases the
GRB signal was too low for any reasonable spectral fit.  For just over
half of the bursts (16) we used the High Energy Resolution Burst
(HERB) data from the most brightly illuminated BATSE LAD (128 energy
channels).  For short bursts or in those cases with no good time
coverage with the HERB data we analyzed the Medium Energy Resolution
(MER) data (5 cases), which consist of 16-energy channel spectra with
a time resolution of 16 ms during the first half of the total duration
and 64 ms during the second half.

The background spectrum was extracted and averaged from two time
intervals, before and after the burst trigger.  The time interval for
the spectrum accumulation typically starts at the trigger and lasts
for the duration as estimated by $T_{90}$.  The spectrum was
eventually re-binned in energy in order to have enough statistics to
apply the $\chi^{2}$ goodness-of-fit test .\footnote{We binned in
order to obtain a minimum of 5 counts/bin. This (minimal) choice has
been tested by re-binning the data for a minimum of 10 and 20
counts/bin (e.g. Bevington \& Robinson 1992) and the corresponding
fits did not yield either significantly different $\chi^2$ nor
significantly different fit parameters.}

The spectra were analyzed using the XSPEC v.11.1.0 package, performing
the fitting with the Band model in the $\sim$30 -- $\sim$1700 keV
energy range.  In some GRBs this model fit resulted in unconstrained
spectral parameters and/or unacceptable reduced $\chi^{2}$ (18
cases): if mainly due to the low statistics at high energies we tested
as alternative models the cutoff-power law (CPL; for 1 GRB) and broken
power--law (BPL; 6 GRBs) spectral functions, while if due to a low
signal in most channels, we adopted the simpler single power-law model
fit (11 GRBs).  The same set of models was applied to larger samples of
bright bursts (Preece et al. 2000): we stress that our aim is to
characterize the spectra of the GRB/SN sample rather than establish
which model preferentially fits them (in fact the choice of a
particular model may not be unique even for bright bursts, e.g.
Ghirlanda et al.  2002).  We considered the model fit acceptable when
a reasonable $\chi^{2}$ (ranging from 0.8 to 1.3) and non systematic
residuals were obtained. All the models (except for the single
power--law) are characterized by a low energy power--law $\propto
E^{\alpha}$, a high energy power--law (Band and BPL models) $\propto
E^{\beta}$ and a characteristic energy $E_{\rm peak}$.

\subsection{Results}
\begin{figure}
\centerline{\epsfig{figure=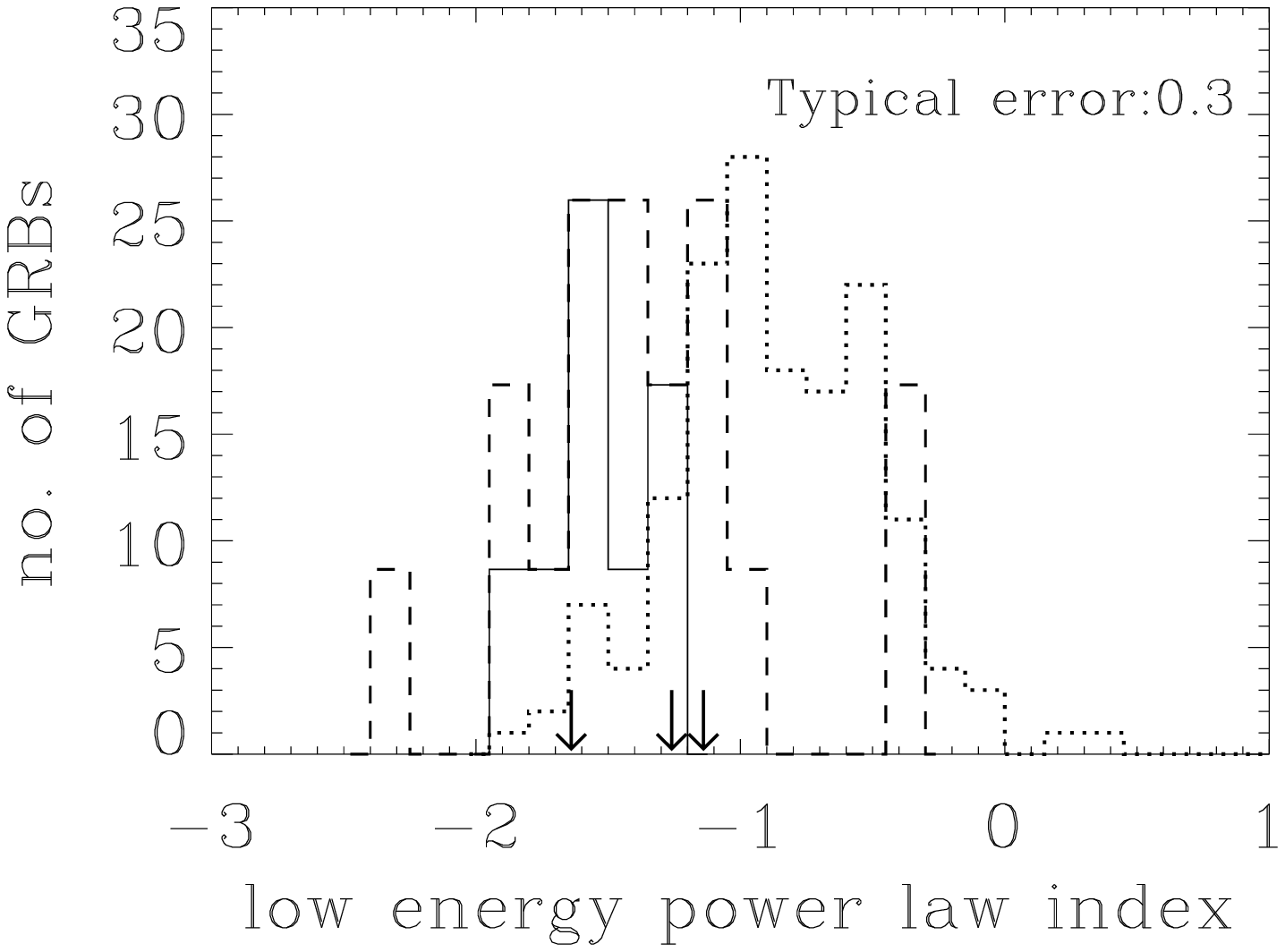,width=8.0cm,height=4.8cm}}
\centerline{\epsfig{figure=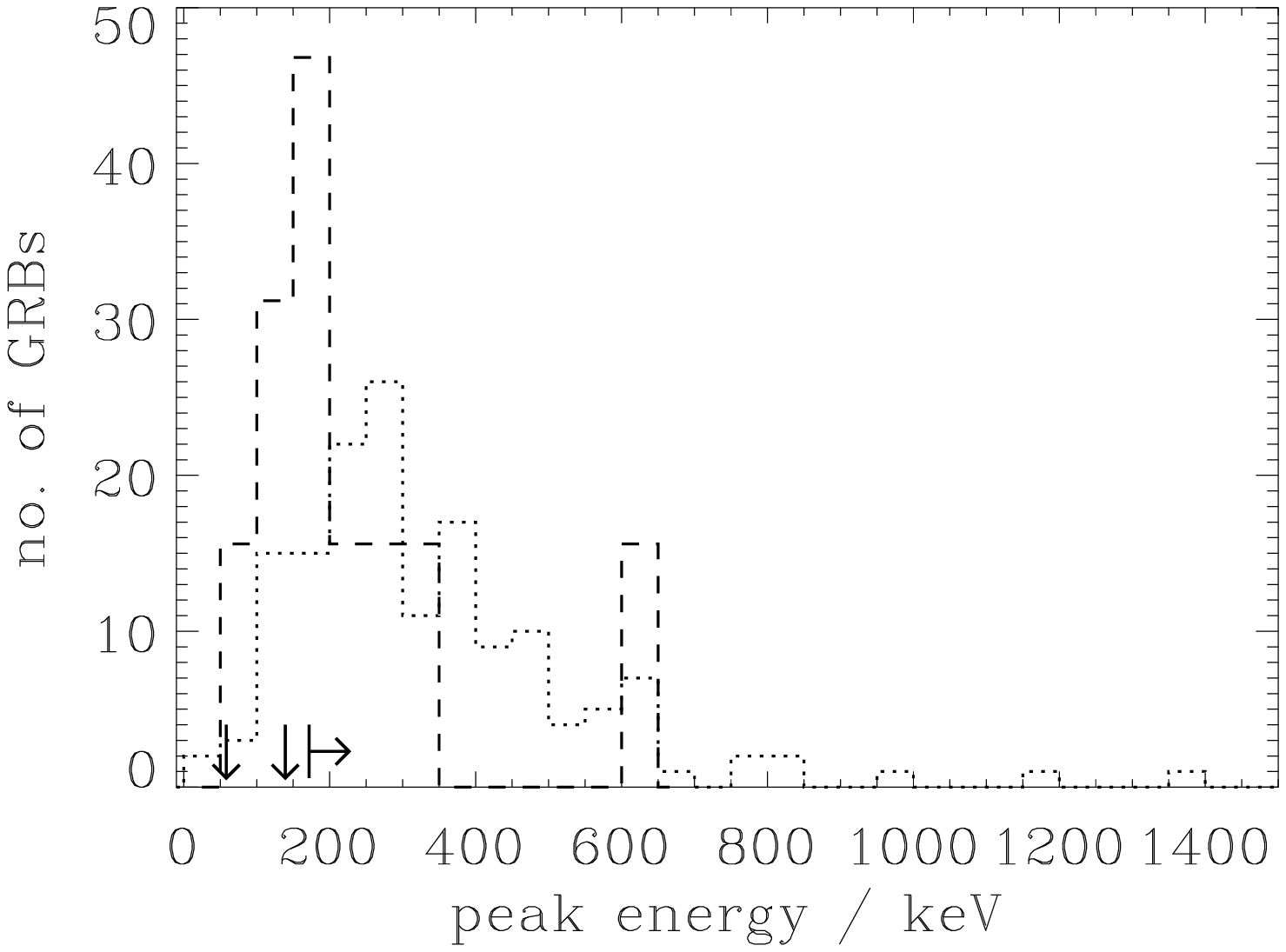,width=8.0cm,height=4.8cm}}
\centerline{\epsfig{figure=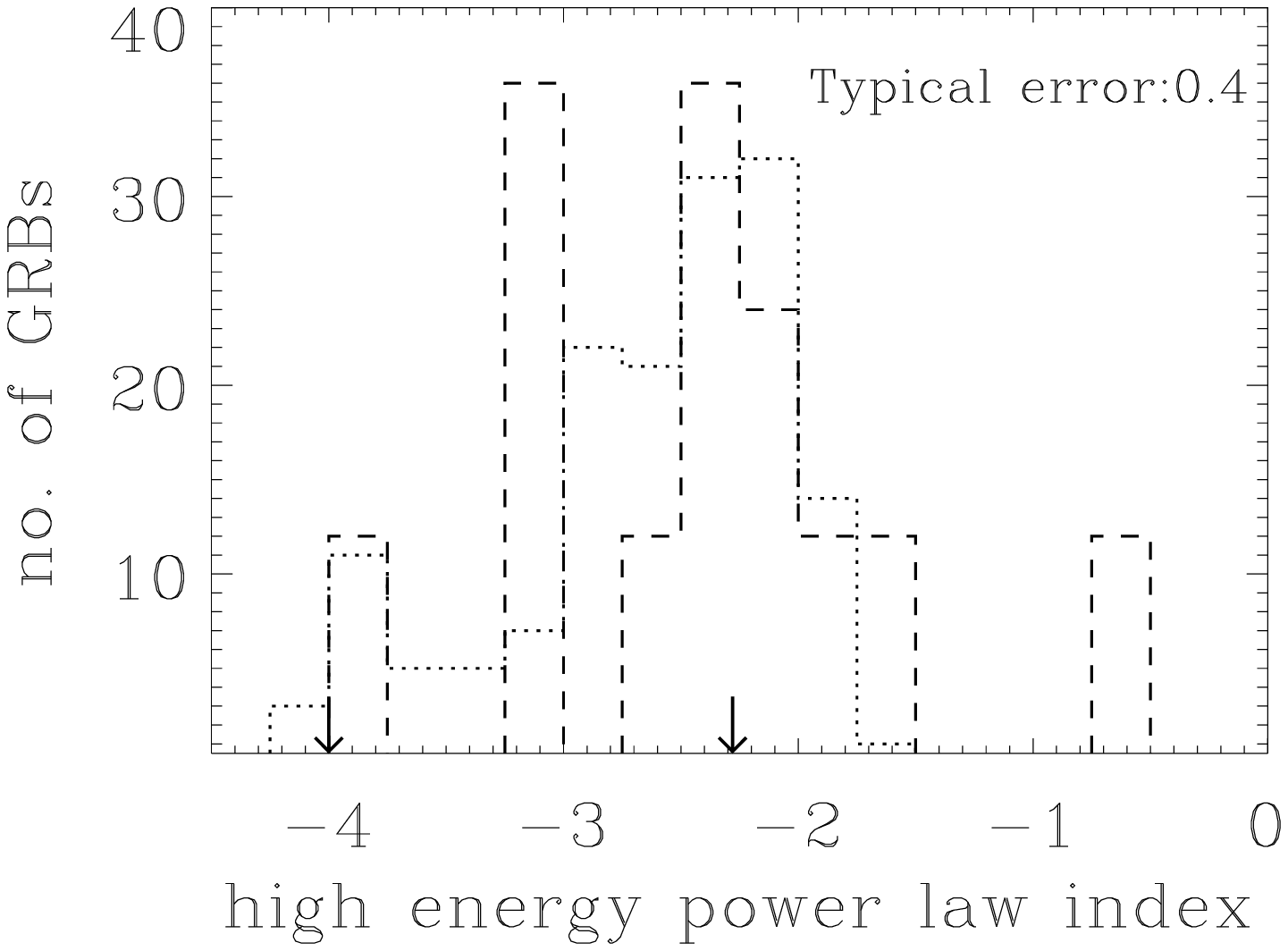,width=8.0cm,height=4.8cm}}
\caption{Spectral parameter distributions obtained for the GRB/SN
(dashed lines) and GRBs examined by Preece et al. (2000) (dotted
lines). For clarity the number of GRB/SN events is normalized to the
total number of GRBs in the larger control sample by Preece et
al. (2000).  In the top panel the results obtained by fitting a single
power law model are indicated separately (solid line). A high energy
power--law index $\beta=-4$ refers to the exponential function
fit. Errors for the peak energy range from 20 to $\sim$ a few hundred
keV. The arrows in each panel indicate the values (or limits)
corresponding to GRB~980425, GRB~030329 and GRB~031203. Note that the
highest values in the three distributions refer to GRB~990810 and
GRB~971221, found by catalog cross--correlation.}
\end{figure}
 
The distributions of the spectral parameters are shown in Fig.~7 for
the GRB/SN and the Preece et al. (2000)'s sample
\footnote{For the comparison the results of the time-resolved spectral
analysis by Preece et al.  were averaged, as for most GRB/SN it was
possible to analyze only time integrated spectra.}. 

For the GRB/SN sample the average spectral index is $\alpha
\sim$--1.5, with almost no bursts with $-1<\alpha<0$, which are
instead found in the comparison sample.  This suggests that on average
GRB/SN events have low energy spectra softer than typical GRBs.
Note that GRB 980425 and GRB 030329 have spectra fitted with $\alpha
\sim$--1.2 (Vanderspek et al.  2004) and GRB 031203 is modeled with a
single power law with $\alpha \sim$--1.6 (Sazonov et al. 2004).  
It is important to note that the single power law model (fitted to
$\sim$ half of GRBs) in general gives a value of $\alpha$ lower than
that inferred from `convex' models (Band, CPL or BPL).  A K-S test
performed on the $\alpha$ values obtained by fitting the `convex'
models only yields that the GRB/SN sample is significantly different
from that reported by Preece et al. (2000), with P$_{KS}$=0.03.

The distributions of $E_{\rm peak}$ (for the 10 GRB/SN events that had
spectrum fitted by Band, CPL or BPL) are reported in the middle panel
of Fig.~7.  The typical peak energy for GRB/SN is $\sim$ 220 keV,  lower 
than that of bright BATSE GRBs (from Preece et al. 2000) 
 with P$_{KS}$=0.04. As mentioned, 2 out of 3 GRBs with spectroscopic 
evidence of association lie in the lowest bin of this 
distribution (see Fig.~7).

Finally we report in Fig.~7 (bottom panel) the distributions of
$\beta$. Within the large errors ($\beta$ is usually not well
constrained because of the S/N decrease in the high energy channels),
there is no evidence of deviations with respect to bright long bursts.

\section{Redshift estimates}  
\begin{figure}
\centerline{\psfig{figure=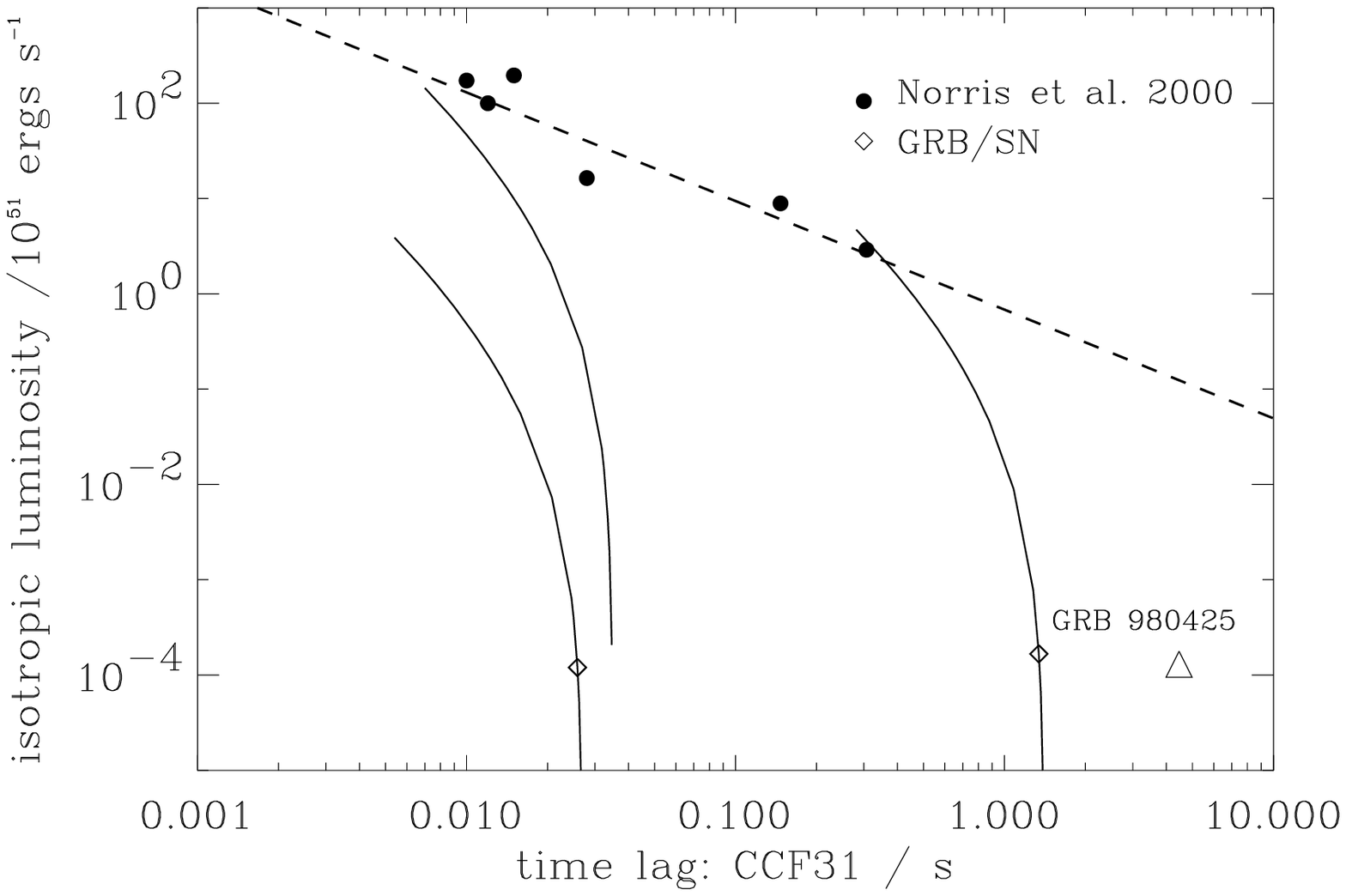,width=8.1cm}}
\centerline{\psfig{figure=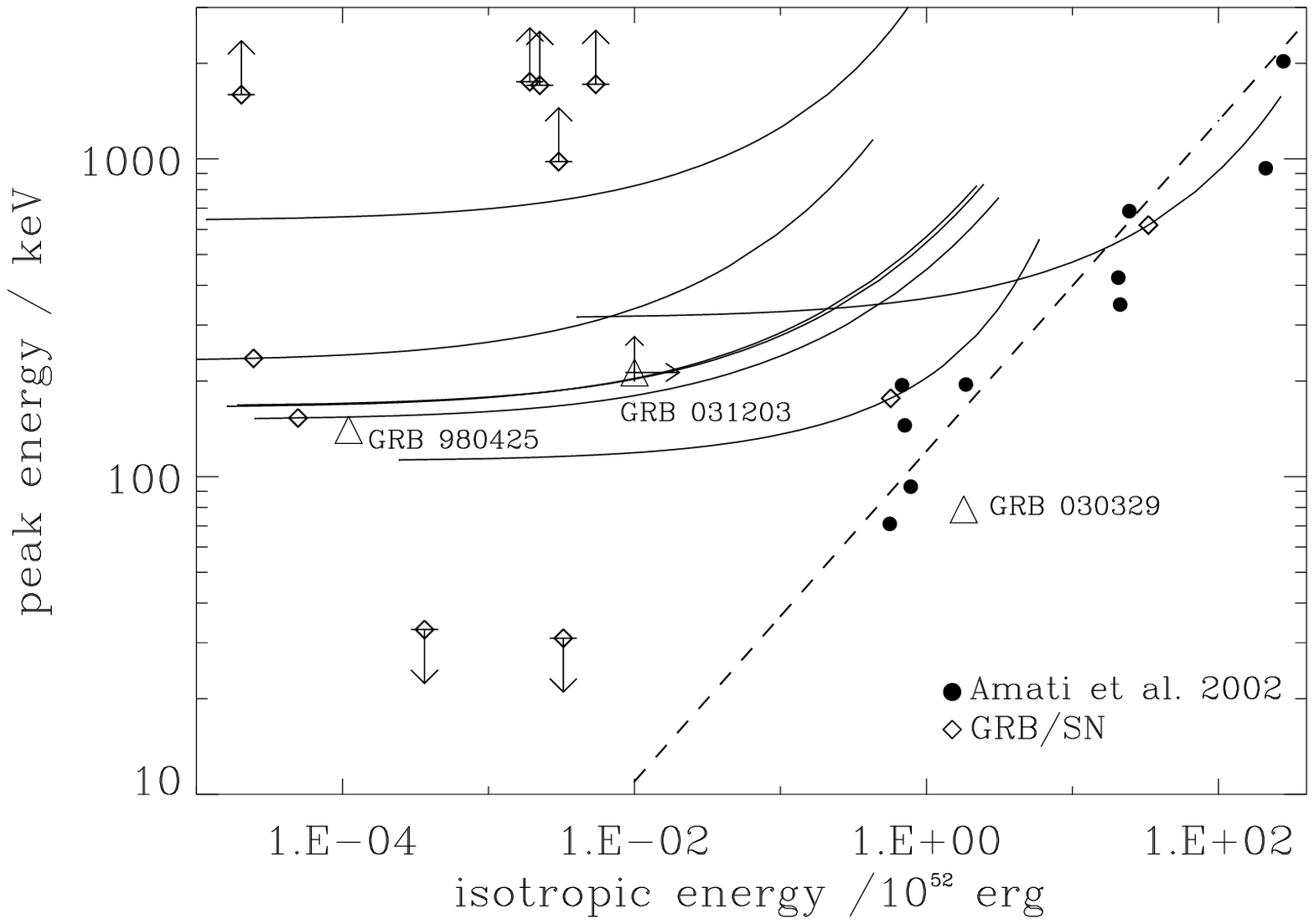,width=8.1cm}}
\caption{Top panel: Relation by Norris et al. (2000).  Bottom panel:
Relation by Amati et al. (2000).  In both panels, the GRB/SN events
are marked (diamonds) at the values corresponding to the redshift of
the associated SN, while the trajectories represent the values
corresponding to different redshifts, in the range 0.01-4.  The dashed
lines indicate the correlations as determined by Norris et al. (2000)
and Amati et al. (2002).  GRB/SN events which had only upper or lower
limits on their peak energy are shown only for the values
corresponding to the redshift of the coincident SN as arrows.}
\end{figure}

Clearly a crucial test on the physical connection of our GRB/SN pairs
would be the determination of the GRB redshifts.  Up to date,
redshifts are known only for $\sim$ 42 GRBs via the optical emission
lines of their host galaxies or absorption lines in the optical
transient spectra.  Due to the high $\gamma$-ray flux, unaffected by
extinction, and to the limited number of afterglow detections, several
attempts have been made in order to find a reliable method for
deriving distances from the prompt $\gamma$--ray emission
only. Unfortunately the large diversity in the temporal and spectral
properties of GRBs and the related uncertainties in turn result in
significant uncertainties in any such method.  For our purposes
however even an indication of the GRBs distance could rule out or
reinforce an association.

Several correlations have been proposed to infer the GRB energetics on
the basis of observable temporal and spectral $\gamma$-ray properties.
Fenimore \& Ramirez-Ruiz (2000) (see also Reichart et al.  2001) found
a correlation between the luminosity and a variability indicator for a
few GRBs with measured $z$: smoother bursts appear to be intrinsically
less luminous.  Another distance indicator was identified by Norris et
al.  (2000) as a correlation between GRB luminosities and spectral
lags.  These correlations have then been used to estimate redshifts
for a large sample of GRBs (Fenimore \& Ramirez-Ruiz 2000).  Another
interesting correlation has been found by Amati et al. (2002) (see
also Lloyd-Ronning \& Ramirez-Ruiz 2002) between the intrinsic peak
energy $E_{\rm peak}$ and the isotropic total radiated energy $E_{\rm
rad}$. Alternative methods to determine $z$ have been proposed: Atteia
(2003) identified a method by combining the spectral parameters and
GRB duration; Bagoly et al. (2003) suggested a correlation between the
peak flux ratios and $z$.

It is worth noticing that the calibration of all these relations is
based on GRBs with known $z$, but always excluding GRB~980425 due to
its peculiar energetics and temporal/spectral properties.  Also
GRB~031203 turned out to be an outsider of some of those correlations.
Caution should be thus considered in using the above results for
candidate GRB/SN events, which might indeed be most similar to
GRB~980425 and GRB~031203.

In the following we focus on two of such methods, for which we could
infer more reliably the corresponding GRB parameters, namely the
Lag-Luminosity correlation (Norris et al. 2000) and the Peak
Energy--Energy  correlation (Amati et al. 2002).

\subsection{Lag-Luminosity correlation}

Norris et al.  (2000) found that the spectral time lag $\tau$ is
anti-correlated with the isotropic peak luminosity $L_{\rm iso}$,
namely L$_{\rm iso, 53}\approx1.3\times(\tau$/0.01 s)$^{-1.15}$ for a
sample of 6 bursts with known $z$.  In order to derive the lag measure
for the GRB/SN the same methodology was applied.  In particular: i) we
calculated the cross correlation functions (CCFs) of the time histories
in two different energy channels (see also Band 1997), i.e. CH1 and
CH3, adding the apodization intervals before and after the burst
(apodization intervals had a constant count number, computed as
average value in adjacent time bins); ii) the analysis was limited to
light curve regions with count rates larger than 10\% of the peak
luminosity; iii) the time series were binned in order to have higher
S/N (64 ms data were binned to 128 ms; for short GRBs TTE data were
binned to 4 ms); iv) we fitted a linear background to every channel
separately in the non-burst portions of the data and subtracted it
from the total counts; v) finally the CCFs were fitted with a third
order polynomial around their peaks in order to derive an estimate of
the time lag, then corrected by 
\footnote{Note that both time dilation and channel energies should be
corrected for $z$ (see e.g.  Fenimore $\&$ Ramirez-Ruiz 2000).} a
factor 1/(1+$z$).  We derived the peak luminosity in the 50-300 keV
energy range directly by integration of the GRB fitted spectrum for
each possible redshift.

The results are presented in Fig.~8 (top panel). Unfortunately it was
possible to estimate time lags for only 3 bursts (16 GRB/SN have both
spectral data and four-channels time histories available and of those
most have negligible signal in CH1; also in 3 cases the obtained time
lags were negative and thus not further considered). The curves
represent the intrinsic (i.e. corrected for $z$) $L_{\rm iso}$ and
$\tau$ assuming a range $z$ = 0.01 - 4 for the possible redshift.  The
points on them indicate the values of $L_{\rm iso}$--$\tau$
corresponding to the redshift of the putative associated SN, when
known. It appears that the GRB/SN events could be consistent with the
Lag-Luminosity correlation only if they were placed at high $z$
($>$1.5) indicating that the association are spurious. Alternatively,
these GRBs might not obey such correlation and thus could still be
located at the SN redshift.  The latter (intriguing) possibility does
in fact get some credit from the fact that these GRBs would then mimics
the characteristics of 2 out of the 3 spectroscopic GRB/SN
(i.e. GRB~980425 and GRB~031203), possibly forming a separate class.

\subsection{Peak Energy--Energy correlation}

The Peak Energy--Isotropic Energy correlation (Amati et al. 2002) has
the form $E_{\rm peak} \propto$ $E_{\rm rad}^{0.52}$.  Similarly to
the procedure described above we computed for the GRB/SN the total
bolometric fluence (see Amati et al.  2002) by integrating (1-10$^4$
keV) the best fit spectrum (11 GRB/SN events).  The results are
presented in the $E_{\rm peak}$--$E_{\rm rad}$ plane in Fig.~8 (bottom
panel). The position of the symbols again indicates the corresponding
values for the SN redshift, while the curved lines show the values for
the interval $z$=0.01-4.  The GRB/SN events for which only lower/upper
limits of the peak energy could be estimated are shown only at the $z$
of the possibly associated SN.  We also mark the values for the 3
spectroscopic associations.  Only two of the GRB/SN could be
consistent with the Amati et al.'s relation: GRB~980703 (at $z$=0.967,
Djorgovski et al. 1998), and GRB~971221 which might be associated with
a SN at $z\approx0.58$ (see Table~1).  The properties of the rest of
GRB/SN are not consistent with those predicted by such relation, but
most importantly in 10 cases (5 of which with well determined
parameters) could not obey the correlation (within its scatter) for
{\it any} $z$.  As known GRB~980425 and GRB~031203 are also clear
outliers of the Amati's relation (whereas GRB~030329 is consistent
with it), i.e. of the strong associations 2/3 do not follow it
(with respect to about 40 GRBs with known redshift).

To conclude, within the limitations given by the (small) fraction of
the GRB/SN sample we interestingly find that these events do not
follow the Lag-Luminosity and the Peak Energy-Isotropic Energy
correlations, in some cases for any assumed $z$.  This -- supported by
the similarity with the behavior of GRB~980425 and GRB~031203 --
appears to indicate that at least a few GRB/SN events form a separate
class with respect to the majority of long GRBs.

\section{Summary and conclusions}

We presented a detailed analysis of the temporal and spectral
properties of a sample of BATSE GRBs with an indication to be
associated with a SN event, with the aim of determining whether these
reveal any peculiar property with respect to the bulk of the detected
GRBs.

The ``GRB/SN sample'' was obtained combining GRB associations based on
the presence of re-brightening in the afterglow light curves which
might be interpreted as SN signatures and (more tentative) on
positional/temporal coincidence from the cross correlation of GRB and
SN catalogs.  The latter ensemble was expanded with respect to
previous works, by including the most recent and complete GRB and SN
catalogs. The whole set comprises 36 GRBs (3 of which lack a
corresponding SN).  For GRBs with public data we performed a systematic
analysis and compared their temporal and spectral properties with
those of larger, ``control'' samples of BATSE bursts.  

The analysis of the temporal profiles revealed that in most cases the
light curve presents a {\it single peak}.  This property has been
already considered by different authors (also on theoretical grounds)
as a possible distinctive feature of GRBs associated with SN events
(Bloom et al. 1998, Norris et al. 1999).  The chance probability of
finding the same fraction of single peaked bursts from a random
extraction from the BATSE archive turned out to be $\sim$ 0.4 \%.

We further analyzed the time profiles fitting a lognormal model and
correlated the pulse parameters. Dim bursts turned out to have {\it
time lags} larger than bright ones, as possibly explained by a larger
viewing angle (Ioka \& Nakamura 2001).  This interpretation has also
been invoked in the case of GRB~980425/SN~1998bw and
GRB~031203/SN~2003lw to account for their low luminosity compared to
the bulk of GRBs (Waxman 2004).

In terms of the spectral properties the GRB/SN sample shows {\it
softer} spectra (both in terms of low energy spectral index and peak
energy) with respect to typical bursts.  This might be a distinctive
feature of GRBs associated with SN as 2 out of 3 spectroscopic
associations also share this atypical low peak energy.

Finally, in the attempt to estimate their redshift, we considered two
of the intrinsic correlations proposed for the population of long
bursts, namely the Lag--Luminosity (Norris et al. 2000) and the
$E_{\rm peak}$--$E_{\rm rad}$ (Amati et al.  2002) relations.  While
some GRB/SN might be consistent with them if not associated to the
putative SN, some events do not obey such correlations for {\it any}
redshift. This might indicate that they are effectively different or
that these correlations only hold for particular sub-samples of GRBs
(e.g. bright long bursts - see also Ghirlanda et al. 2004 who find
that short bursts mostly populate a region above the $E_{\rm
peak}$--$E_{\rm rad}$ correlation).  In particular, the properties of
this handful of GRB/SN again closely resemble those of 2 out of 3
spectroscopic associations which are also outliers of those relations.

As discussed in the Introduction, previous results as well as the
large positional uncertainties and wide time window for the selection
of GRB/SN via catalog cross--correlation in fact point toward a low
probability of finding true associations via this method.  However,
the above results are suggestive of the opposite.  Statistically, the
GRB/SN sample contains (with respect to the whole of the BASTE events)
an unusually high fraction of single peaked bursts, with softer
spectra and some of those do not follow two of the proposed
correlations for GRBs. Most of these properties are mimicking those of
the spectroscopic GRB/SN associations.  Although it is not obvious how
to classify individual GRB/SN, in order to see any intersection among
GRB/SN having `unusual' characteristics, in Table~1 we schematically
listed the three `unusual' properties for each GRB/SN
(single--peakness, spectrum and consistency with the $E_{\rm
peak}$--$E_{\rm rad}$ relation).  In 19 GRBs (other than GRB~980425)
there are at least 2 properties which set the event as
GRB~980425-like: 10 GRB/SN have 2 or 3 (3 of those are short and only
1 not selected by the catalog cross-correlation); 6 GRBs have 2 but not
the third one (5 are short and 2 not selected by the catalog
cross-correlation).

A final indication that the catalog cross--correlations are not
randomly sampling GRB and SN events comes from the properties of the
selected SNe.  We did not restrict our search for correlations on the
basis of the SN Type, yielding a significant number of SNe Ia in the
sample (see Table 1).  While indeed it has been also proposed on
theoretical grounds that also Type Ia SN might be associated to bursts
(e.g. Berezinsky et al. 1996; Dar \& De Rujula 2004), the current view
(theoretically and observationally based) is that (some long) GRBs are
likely associated with Type Ibc SNe. The intriguing finding is that
among the GRB/SN events found by the catalog cross-correlations Type
Ib,c SNe comprise about 28\% of the events (i.e. 8/29), which is about
4 times larger than expected from SN catalog statistic (see Valenti et
al. 2005 who examined the Asiago SN catalog). This result also
confirms the analysis independently carried out by Valenti et al. 2005
on BATSE catalog; indeed, these authors found an excess of Ib,c Types
among SNe possibly associated with BATSE GRBs ($\sim$17\%, with a
significance level of $\sim$ 93\%).

While the uncertainty on the reality of a single association remains
high, these results indicate that: 1) there is a population
significantly larger than that detected so far of GRB 980425-like
events associated with SN; 2) the corresponding GRB have ``atypical''
temporal and spectral properties (and thus may suggest that not all
GRBs are associated with at least typical `SN'); 3) they appear
under-luminous in their observed $\gamma$--ray emission (at the level
of 0.1-1 \%) with respect to ``standard'' GRBs.

All three aspects give statistical and physical information on the
nature of GRB 980425-like events.  The finding of a significant number
of these under-luminous nearby bursts (e.g. viewed off axis) might in
fact indicate from statistical arguments that these comprise the
majority of GRBs (e.g. Sazonov et al. 2004).  From the physical point
of view, one promising scenario accounting for a qualitatively softer
spectrum, a smooth, single--peak light curve and an apparently
under-energetic event could be envisaged if the line of sight is at
large angles with respect to the jet axis and the observed GRB
emission is in fact prompt emission reprocessed (reflected) at such
angles by circumburst material. A detailed study of the spectral and
temporal properties expected in such case is ongoing.

Estimates of the bolometric energetics of GRBs (e.g.  Soderberg 2004)
and the improved statistics which will be allowed in the HETE,
INTEGRAL and starting Swift era offer the most obvious possibilities
to test these results.

\begin{acknowledgements}
We thank the referee, E. Cappellaro, for useful criticisms and
valuable suggestions. ZB thanks T. Maccarone for helpful comments. EP
thanks P. Mazzali and M. Vietri for the initial discussions on this
project.  The Italian MIUR and INAF are acknowledged for financial
support by ZB, AC and GG (MIUR/COFIN Grant 2003020775-002).  This
research was supported in part by the National Science Foundation
under Grant No. PHY99-07949; the KITP (Santa Barbara) is thanked for
kind hospitality (AC). The research has made use of data obtained
through the High Energy Astrophysics Science Archive Research Center
online service, provided by the NASA/GODDARD Space Flight Center.

\end{acknowledgements}

\newpage
\begin{table*}
{\tiny \begin{tabular}{cccccccccccccccccc} \hline\noalign{\smallskip}
GRB & RA & DEC & Error & T$_{90}$ & $P_{\rm 64 ms}$ & $F_{\rm >25
keV}$ &Single &$\alpha$ &SN & RA & DEC& Discovery & Type & z$^{@}$ &
Follow & Catalog &Ref. \\ \noalign{\smallskip} & & & box (deg)&
s & ph/cm$^2$s & 10$^{-6}$erg/cm$^2$ &peak$^{*}$ && & && date &&
&$E_{peak}-E_{\rm rad}$ &cross-corr.  & \\ \noalign{\smallskip} \hline
\hline
\noalign{\smallskip} 
 920321 & 184.5 & 5.7 & 3.6 & - & - & - &N&$-1.06^{+0.14}_{-0.13}$   & 1992Q & 182.0 & -1.6 & 920407 &- &- & N& Y&2 \\\noalign{\smallskip} 
 920613 & 312.9 & -55.6 & 4.6 & 129.0 & 1.02 & 1.86 &N&-& 1992ae & 322.1 & -61.6 & 920627 & Ia & 0.075 & N &Y& 1 \\
\noalign{\smallskip} 
920708 & 308.3 & -49.9 & 3.3 & 3.2 &- &- &Y$^{*}$&  $-1.41^{+1.54}_{-0.17}$& 1992al & 311.5 & -51.4 & 920727 & Ia & 0.014 &N&Y& 1 \\
\noalign{\smallskip} 
920628 & 317.8 & -27.3 & 7.7 & 4.5 & 0.52 & 1.63 &Y&-& 1992at & 321.8 & -37.0 & 920717 &Ia &- & - & Y& 1\\ 
\noalign{\smallskip} 
920925 & 129.7 & -58.7 & 5.0 & - &- &- &Y&-& 1992bg & 115.5 & -62.5 & 921016 & Ia & 0.036 &-&Y& 1 \\
\noalign{\smallskip} 
951107 & 148.8 & 39.8 & 4.0 & 43.5 & 0.69 & 1.45 &Y$^{*}$&  $-1.62_{-0.10}^{-0.11}$ &  1995bc & 147.7 & 40.3 & 951201 & II &0.048 &N&Y& 2 \\ 
\noalign{\smallskip} 
960221 & 47.8 & -31.2 & 4.3 & 31.3 & 0.71 & 4.22 &Y&-& 1996N & 54.7 & -26.3 & 960310 & Ib&-  &-&Y& 1\\
\noalign{\smallskip} 
961029 & 59.3 & -52.6 & 3.3 & 40.4 & 1.03 & 8.44 & Y&-& 1996bx & 59.7 & -53.3 & 961118 & Ia & 0.069 &-&Y& 2 \\
\noalign{\smallskip}
970508 & 132.4 & 80.6 & 2.3 & 23.1  &1.28 &3.96 &Y& $-1.46_{-0.23}^{+0.10}$ &- & -&- &- &- &0.835 &Y$^{+}$&N& 3 \\
\noalign{\smallskip} 
970514 & 67.6 & -60.9 & 3.7 & 1.3 & 4.83 & 0.41 &Y&-& 1997cy & 68.0 & -61.7 & 970716 & II & 0.0642 &-&N& 4 \\
\noalign{\smallskip}
971218 & 116.1 &  16.7 & 5.2 & 6.8 & 0.76 & 1.09 & Y& $-1.25_{-0.58}^{+0.31}$  &1998B & 116.5 & 18.7 & 980101 & Ia & 0.045&N&Y & \\
\noalign{\smallskip}
971221 & 73.7 & 4.7 & 6.3 & 1.0 & 2.87 & 0.58 &Y&$-2.35^{+0.57}_{-1.56}$ & 1997ey & 74.2 & -2.6 & 971229 & Ia & 0.58 &Y&Y& \\
\noalign{\smallskip}
980326 & 133.3 & -18.6 & 2.1&-&-& -&   Y$^{*}$& $-1.90_{-0.20}^{+0.26}$ &-&-&-&- &- &0.9--1.1 &Y$^{+}$&N& 5 \\
\noalign{\smallskip} 
980425 & 291.9 & -53.1 &1.6 & 34.8 & 1.24 & 4.0 &Y& $-1.14_{-0.21}^{+0.22}$  & 1998bw & 293.7 & -52.8 & 980428 & Ic&0.0085 &N&N&  6 \\
\noalign{\smallskip}
980525 & 157.6 & -17.8 & 6.8 & 39.6 & 1.40 & 0.38 &N& $-1.69_{-0.19}^{+0.18}$  & 1998ce & 152.6 & -25.8 & 980519 & II&- &-&Y& \\
\noalign{\smallskip}
980703 & 359.1 & 12.0 & 0.5 &411.6 & 2.93 & 62.2&N& $-1.01_{-0.12}^{+0.10}$ &- &- &- &- &-&0.967 &Y&N& 7 \\
\noalign{\smallskip}
980910 & 195.1 & -21.1 & 6.8 &- &- &- &-&- &1999E & 199.3 & -18.5 & 990115 & IIn&0.025 &-&N& 8 \\  
\noalign{\smallskip}
990527 & 199.9 & 49.3 & 7.7 &18.9 & 0.58 & 1.04 &Y &$-1.65_{-0.40}^{+0.33}$ &1999ct  &198.3 &46.2 &990613 &Ia &0.18 &-&Y& \\ 
\noalign{\smallskip}
990810 & 358.1 & 1.7 & 4.3 & 0.1 & 4.63 & 0.70&Y& $-0.32_{-0.40}^{+0.40}$ & &multiple &(2) &associations & &     &N&Y& \\
\noalign{\smallskip}
991002 & 25.1 & 3.7 & 2.2 &1.9 & 10.2 & 0.62&Y&$-1.12_{-0.26}^{+0.49}$ &1999eb & 25.7 & 3.7 & 991002 & IIn &- &N&N& 9 \\
\noalign{\smallskip}
991015 & 11.6 & 8.9 & 7.8 & 2.7 & 0.91 & 0.47 &Y& - & 1999ef & 14.7 & 12.7 & 991009 & Ia &- &-&Y& \\
\noalign{\smallskip}
991123 & 157.2 & 26.9 & 4.7 & - & 2.95 & 0.31&Y&- & 1999gj & 10.5 & 26.1 & 991117 & Ia & 0.018 &-&Y& \\
\noalign{\smallskip}
000114 & 121.6 & 39.5 & 7.9 &0.58 & 1.35 & 0.12&Y& $-1.42_{-0.21}^{+0.21}$ & 2000C & 114.2 & 35.2 & 000108 & Ic&0.012 &N&Y& \\
\noalign{\smallskip}
000415 & 199.4 & -29.9 & 7.4 & 0.22 & 2.72 & 0.40&Y& $-0.39_{-0.49}^{+0.74}$ & 2000ca & 203.8 & -34.1 & 000428 & Ia& 0.024 &N&Y&\\
\noalign{\smallskip}

\hline\noalign{\smallskip}  
910423 & 196.7 & -5.2 & 11.1 & 208.5 & 0.41 & 0.25 &Y &$ -1.88^{+0.40}_{-0.51} $& 1991aa & 190.5 & -6.0 &910507 & Ib &- &-&Y&  1 \\
\noalign{\smallskip}
950917 & 339.9 & -1.47 & 8.2 &- &- &- &Y&- & 1995ac & 341.4 & -8.8 & 950926 & Ia & 0.05 &-&Y& 2  \\ 
\noalign{\smallskip}
960925 & 29.3 & -13.9 & 8.4 &1.8 & 1.02 & 0.47 &Y&- &  1996at & 17.1 & -1.0 & 961009 & Ib/c & 0.09 &-&Y& 2\\
\noalign{\smallskip}
961218 & 97.7 & -21.7 & 12.7 &8.7 & 0.40 & 0.91 &Y$^{*}$&- & 1997B & 88.3 & -17.9 & 970104 & Ic & 0.01 &N&Y&2 \\
\noalign{\smallskip}
970907 & 346.9 & 11.2 & 8.3 &0.9 & 7.24 & 2.35 & Y& $-1.26_{-0.16}^{+0.14}$ & 1997dg & 355.1 & 26.2 & 970928 & Ia &0.03 &N&Y&2  \\
\noalign{\smallskip}
971013 & 167.0 & 2.7 & 8.8 & 12.3 &- &- &Y&- & 1997dq & 175.2 & 11.5 & 971105 & Ib &0.003 &-&Y& 2 \\
\noalign{\smallskip}
971120 & 155.7 & 76.4 & 9.9 & 2.2 & 0.71 & 0.77 &Y&- & 1997ei & 178.5 & 58.5 & 971223 & Ic &0.011 &-&Y&2 \\
\noalign{\smallskip}
980530 & 148.3 & -27.3 &8.7 & 21.3 & 0.67 & 0.79 &Y&$-1.51_{-0.15}^{+0.14}$ & 1998ck & 145.2 & -29.1 & 980531 
& Ia & 0.038 &N&Y&\\
\noalign{\smallskip}
990719 & 224.5 & 10.8 & 10.9  & 23.2 & 0.82 & 0.23 &N&- & 1999dg & 227.8 & 13.5 & 990723 & Ia &- &-&Y& \\
\noalign{\smallskip}
990902 & 58.2 & 62.5 & 15.5 &19.2 & 0.42 & 0.85 &Y&- &   1999dp & 67.3 & 69.5 & 990902 & II &0.016 &-&Y&\\
\noalign{\smallskip}
991021 & 346.3 & -30.9 & 12.3 & - & 1.30 & 0.07&Y&- &  1999ex & 334.0 & -36.8 & 991109 & Ic &0.011 &-&Y&\\
\noalign{\smallskip} 
000319 & 172.7 & -13.8 & 9.8 &0.08 & 1.75 & 0.10&Y&- & & multiple& (5)& associations& &-&-&Y&\\
\noalign{\smallskip}
\hline
\end{tabular}}
\caption{The GRB/SN sample. The line separates GRBs with BATSE error
boxes smaller and larger than 8 deg. ($^{*}$) refers to a
single-peaked GRB with substructure.  There is no SN reported when
there was only a SN bump present in the afterglow or in case of
multiple (\#) associations.  $^{@}$ IAUC SN catalog. $^{+}$ Already in the Amati et al. (2002) sample;
$^{1}$ Wang \&
Wheeler (1998); $^{2}$ Hudec et al. (1999); $^{3}$ Piro et al. (1999);
$^{4}$ Germany et al.  (2000); $^{5}$ Bloom et al. (1999); $^{6}$
Galama et al. (1998); $^{7}$ Holland et al. (2001); $^{8}$ Rigon et
al. (2003); $^{9}$ Terlevich et al. (1999).}
\end{table*}

\end{document}